\documentclass[letterpaper,english,reprint, aip]{revtex4-1}
\usepackage[T1]{fontenc}
\usepackage[latin9]{inputenc}
\usepackage{babel}
\usepackage{array}
\usepackage{textcomp}
\usepackage{multirow}
\usepackage{amsmath}
\usepackage{amssymb}
\usepackage{graphicx}
\PassOptionsToPackage{normalem}{ulem}
\usepackage{ulem}
\usepackage[unicode=true]
 {hyperref}

\makeatletter

\pdfpageheight\paperheight
\pdfpagewidth\paperwidth

\providecommand{\tabularnewline}{\\}

\usepackage{hyperref}   % use for hypertext links, including those to external documents and URLs
\usepackage[capitalise]{cleveref}   % use for referencing figures/equations

\makeatother

\begin{document}
\hyphenation{ISSNet}
\hyphenation{CGSchNet}
\hyphenation{SchNet}
\hyphenation{CGnet}
\title{Machine Learning Implicit Solvation for Molecular Dynamics}
\thanks{This article may be downloaded for personal use only. Any other use
requires prior permission of the author and AIP Publishing. This article
appeared in Chen et al., \textit{J. Chem. Phys.} \textbf{155}, 084101
(2021) and may be found at \url{https://doi.org/10.1063/5.0059915}.}
\author{Yaoyi Chen}
\affiliation{Department of Mathematics and Computer Science, Freie Universität,
Berlin, Germany}
\author{Andreas Krämer}
\affiliation{Department of Mathematics and Computer Science, Freie Universität,
Berlin, Germany}
\author{Nicholas E. Charron}
\affiliation{Department of Physics, Rice University, Houston, Texas 77005, USA}
\affiliation{Center for Theoretical Biological Physics, Rice University, Houston,
Texas 77005, USA}
\affiliation{Department of Physics, Freie Universität, Berlin, Germany}
\author{Brooke E. Husic}
\thanks{Corresponding author}
\email{husic@princeton.edu}

\affiliation{Department of Mathematics and Computer Science, Freie Universität,
Berlin, Germany}
\author{Cecilia Clementi}
\thanks{Corresponding author}
\email{cecilia.clementi@fu-berlin.de}

\affiliation{Department of Physics, Freie Universität, Berlin, Germany}
\affiliation{Center for Theoretical Biological Physics, Rice University, Houston,
Texas 77005, USA}
\affiliation{Department of Chemistry, Rice University, Houston, Texas 77005, USA}
\affiliation{Department of Physics, Rice University, Houston, Texas 77005, USA}
\author{Frank Noé}
\thanks{Corresponding author}
\email{frank.noe@fu-berlin.de}

\affiliation{Department of Mathematics and Computer Science, Freie Universität,
Berlin, Germany}
\affiliation{Department of Chemistry, Rice University, Houston, Texas 77005, USA}
\affiliation{Department of Physics, Freie Universität, Berlin, Germany}
\begin{abstract}
Accurate modeling of the solvent environment for biological molecules
is crucial for computational biology and drug design. A popular approach
to achieve long simulation time scales for large system sizes is to
incorporate the effect of the solvent in a mean-field fashion with
implicit solvent models. However, a challenge with existing implicit
solvent models is that they often lack accuracy or certain physical
properties compared to explicit solvent models, as the many-body effects
of the neglected solvent molecules is difficult to model as a mean
field. Here, we leverage machine learning (ML) and multi-scale coarse
graining (CG) in order to learn implicit solvent models that can approximate
the energetic and thermodynamic properties of a given explicit solvent
model with arbitrary accuracy, given enough training data. Following
the previous ML\textendash CG models CGnet and CGSchnet, we introduce
ISSNet, a graph neural network, to model the implicit solvent potential
of mean force. ISSNet can learn from explicit solvent simulation data
and be readily applied to MD simulations. We compare the solute conformational
distributions under different solvation treatments for two peptide
systems. The results indicate that ISSNet models can outperform widely-used
generalized Born and surface area models in reproducing the thermodynamics
of small protein systems with respect to explicit solvent. The success
of this novel method demonstrates the potential benefit of applying
machine learning methods in accurate modeling of solvent effects for
\emph{in silico} research and biomedical applications.
\end{abstract}
\maketitle

\section{\label{sec:Intro}Introduction}

The solvent environment around macromolecules often plays a significant,
sometimes even decisive, role in both the structure and dynamics of
biological systems.\citep{Laage_ChemRev2017,Lewandowski_Science2015,Daniel_PhilosTransRSocLondonSerB2004}
For example, the so-called ``hydrophobic core'', a key structural
element shared by a diverse variety of protein domains strongly influences
protein folding in aqueous solution.\citep{Ball_ChemRev2008,Kalinowska_JMolModel2017}
The solvent also renders the protein structure flexible enough for
functional conformational changes~\citep{Barron_Biochemistry1997}
and mediates interactions among macromolecules for biological processes~\citep{Ball_ChemRev2008,Levy_AnnuRevBiophysBiomolStruct2006}
as well as drug binding.\citealp{Ben-Naim_BiophysChem2002,Persch_AngewChemIntEd2015,Ladbury_ChemBiol1996}

\begin{figure}
\begin{centering}
\includegraphics[width=1\columnwidth]{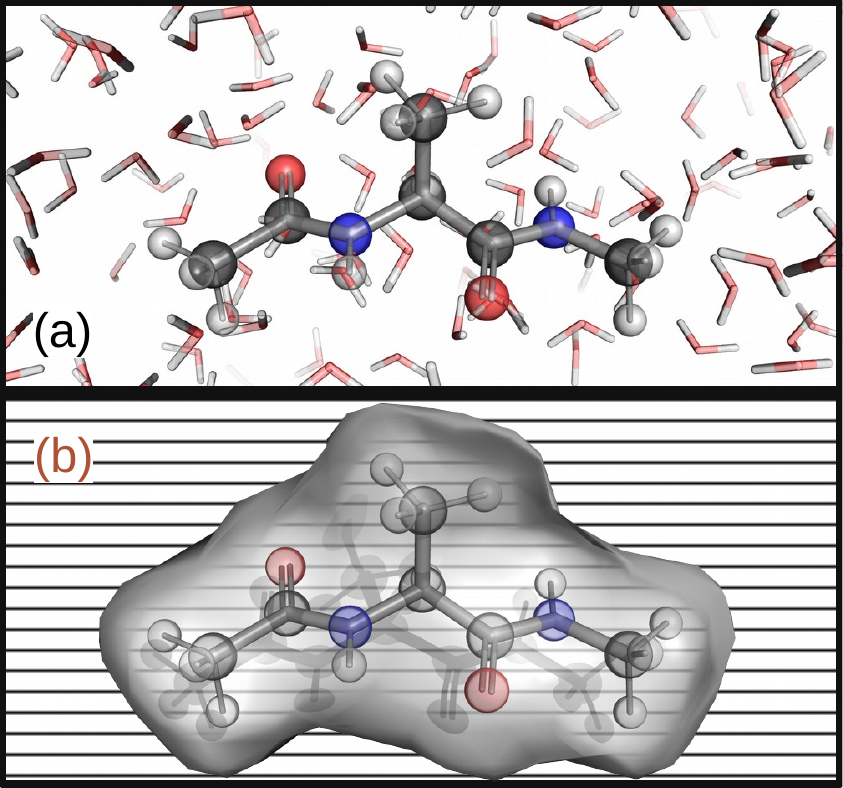}
\par\end{centering}
\caption{\label{fig:ExplImplSolv}(a) Explicit and (b) implicit solvation treatment
of a biomolecular system (here we use capped alanine as an example).}
\end{figure}

Thus, for computational investigations of biomedical problems, such
as molecular dynamics~(MD) simulations of biological systems~\citealp{Adcock_ChemRev2006,Dror_AnnuRevBiophys2012,Braun_LiveCoMS2018}
and molecular docking,\citep{Brooijmans_AnnuRevBiophysBiomolStruct2003}
we often seek to accurately model effects of the solvent environment.
In MD simulations, solvation methods can be grouped into two major
categories: explicit and implicit. The former\textemdash as illustrated
in Fig.~\ref{fig:ExplImplSolv}a\textemdash incorporates solvent
molecules explicitly into the simulation system, while the latter
(see Fig.~\ref{fig:ExplImplSolv}b) represents solvent effects in
a mean-field manner.\citep{Adcock_ChemRev2006,Onufriev_WaterModels2018,Ren_QRevBiophys2012}
Treating the solvent implicitly has several advantages:~it can speed
up force calculations by drastically reducing the the number of degrees
of freedom; it increases the effective time step size in MD simulations;\citep{Anandakrishnan_BiophysJ2015}
and it simplifies constant-pH simulations~\citep{Baptista_JChemPhys2002,Mongan_JComputChem2004}
as well as enhanced sampling approaches, such as parallel tempering~(PT)/replica-exchange
MD.\citep{Trebst_JChemPhys2006,Earl_PhysChemChemPhys2005} Moreover,
implicit solvent treatment is very common in structure-based drug
design, such as fragment screening and lead optimization.\citealp{Sledz_CurrOpinStructBiol2018,Wang_ChemRev2019}
Some generalized Born~(GB)-based implicit solvent methods, for example,
are implemented in various MD software packages, such as GBSA\textendash HCT,\citep{Hawkins_JChemPhys1996}
GBSA\textendash OBC and GBn models~\citep{Roe_JPhysChemB2007} in
AMBER,\citep{Amber20} GBMV~\citep{Lee_JChemPhys2002,Lee_JComputChem2003}
and GBSW models~\citep{Im_JComputChem2003} in CHARMM.\citep{Brooks_JComputChem2009}
Ref.~\onlinecite{Knight_JComputChem2011} gives an comprehensive
comparison of available implicit solvent models.

Despite their advantages, the accuracy of commonly used implicit solvent
models tends to be inadequate in certain applications, such as the
calculation of solvation free energies \citep{Cumberworth_JComputChem2016}
or the recovery of correct conformational distributions for folded
and unfolded states of proteins,\citep{Zhou_PNAS2002,Nymeyer_PNAS2003,Zhou_Proteins2003}
thereby limiting their usage and effectiveness in practice.

The present work addresses a long-standing question in solvent modeling:~is
it possible to construct mean-field implicit solvent models that reproduce
the solvation thermodynamics of explicit-solvent systems exactly?
We approach this problem by parameterizing implicit solvent models
via a machine-learned coarse graining approach. Coarse graining of
molecular systems is itself a well-researched topic, one whose aim
is to model molecules and their interactions with super-atomistic
resolutions, such that computational investigations (e.g., MD simulations)
become more efficient.\citep{Clementi_JMolBiol2000,ClementiCOSB,Saunders_AnnuRevBiophys2013,Noid_JChemPhys2013,Ingolfsson_WIREsComputMolSci2014,Kmiecik_ChemRev2016,Pak_CurrOpinStructBiol2018,Chen_JChemTheoryComput2018,Singh_IntJMolSci2019,Nuske_JChemPhys2019,Wang_ACSCentSci2019,Wang_JChemPhys2020,Husic_JChemPhys2020}
A coarse grained~(CG) model usually entails two important aspects:
the CG resolution and representation\textemdash that is, the mapping
of the original atoms into effective interacting groups (also known
as CG beads)~\citep{Kmiecik_ChemRev2016,Pak_CurrOpinStructBiol2018,Noid_JChemPhys2013,Farrell_JComputPhys2015,Boninsegna_JChemTheoryComput2018}\textemdash and
the CG potential, which determines the interactions among the CG beads.\citep{Pak_CurrOpinStructBiol2018,Noid_JChemPhys2013,Kmiecik_ChemRev2016}
Here we consider an implicit solvent system as a CG version of the
explicit solvent system\textemdash the CG mapping keeps the solute
molecule(s) while removing the solvent degrees of freedom. Once the
CG mapping has been assigned, the parameterization of a CG potential
may follow either a ``top-down'' approach; i.e., one that aims at
reproducing macroscopic experimental observations, or a ``bottom-up''
strategy, which systematically integrates information from the corresponding
atomistic system.\citep{Noid_JChemPhys2013} In this work we leverage
the multi-scale coarse graining theory,\citep{Izvekov_JPhysChemB2005,Noid_JChemPhys2008}
a ``bottom-up'' approach. Essentially, it transforms the parameterization
of a CG potential into a data-driven optimization based on the variational
force matching~(FM) method.

The multi-scale coarse graining theory enables us to employ a machine
learning method similar to the CGnet introduced by Wang et al.~\citep{Wang_ACSCentSci2019}
to learn an implicit solvent model, which is part of the CG potential
for the solute, for any given molecular system. Machine learning methods
have enjoyed increased in popularity and led to breakthroughs in many
fields,\citep{LeCun_Nature2015} including molecular sciences.\citep{Butler_Nature2018,Noe_AnnuRevPhysChem2020,Behler_JChemPhys2016,Chmiela_NatCommun2018}
For structural coarse graining in particular, there have been some
pioneering works both for choosing optimal CG mappings~\citep{Wang_npjComputMater2019,Boninsegna_JChemTheoryComput2018}
and for parameterizing CG potentials for a given system.\citep{Zhang_JChemPhys2018,Wang_ACSCentSci2019,Husic_JChemPhys2020,Wang_npjComputMater2019,Wang_JChemPhys2021}
In this work, we adapt the architecture of CGnet~\citep{Wang_ACSCentSci2019}
and its extension CGSchNet~\citep{Husic_JChemPhys2020} (the latter
based on a graph neural network architecture SchNet~\citep{Schutt_JChemPhys2018})
to the implicit solvent problem. The resulting implicit-solvent SchNet\textemdash henceforth
called ISSNet\textemdash is able to learn an implicit solvent model
from coordinate and force samples of a corresponding explicit solvent
system. Trained ISSNet models can in turn be used for implicit solvent
simulations of biomolecules.

Recently, machine learning methods have been applied in some studies
related to solvent environment, such as the automatable cluster-continuum
modeling of the solvent in quantum chemistry calculations,\citep{Basdogan_JChemTheoryComput2019}
for the parameterization of CG water models for ice-water mixture~\citep{Chan_NatCommun2019}
and liquid water systems,\citep{Patra_ApplPhysLett2019} and for the
computation of generalized Born radii in implicit solvent simulations.\citep{Mahmoud_Bioinformatics2020}
The latter three studies are applicable to MD simulations; however,
the goal is either to achieve higher accuracy for water-only systems
or to improve the efficiency of an existing method. This work distinguishes
itself from existing studies by introducing a neural-network-based
implicit solvent method for biomolecular MD simulations. Additionally,
we are aware of an interesting study which also integrated variational
coarse graining theories to the optimization of an implicit solvent
model.~\citep{Bottaro_JChemTheoryComput2013} However, different
from the relative entropy method used by their work, the multiscale
coarse graining formalism enables simultaneous optimization of all
parameters in a complex neural network model without the necessity
of iterative sampling.

The paper proceeds as follows:~we first describe the theoretical
basis of implicit solvent treatment with ISSNet as well as the implementation,
including the neural network architecture, training and validation
as well as implicit solvent simulation. In the Results we apply our
proposed method to two molecular systems\textemdash capped alanine
(i.e., the solute molecule in Fig.~\ref{fig:ExplImplSolv}) and the
miniprotein chignolin.\citep{Honda_JACS2008} We show that our method
can reproduce the solvated thermodynamics with higher accuracy than
a reference implicit solvent method, namely the GBSA\textendash OBC
model.\citep{Onufriev_Proteins2004} In the Discussion section we
address the current limitation and future investigative directions
of the ISSNet method.

\section{Theory and Methods}

Here, we introduce the potential of mean force~(PMF)\textemdash a
concept from statistical mechanics\textemdash as a theoretical basis
both for implicit solvent methods and for the multi-scale coarse graining
theory. After examining how a traditional approach approximates the
implicit solvent PMF, we adapt an established machine learning CG
method for parameterizing implicit solvent models based on explicit
solvent simulation data.

\subsection{\label{subsec:Theory-SolvPMF}Solute PMF and solvation free energy}

The concept of PMF originated in a 1935 paper by J.~G.~Kirkwood
on statistical treatment of fluid mixtures.\citep{Kirkwood_JChemPhys1935}
In this subsection we derive the expression of a solute PMF following
the framework of Ref.~\onlinecite{Roux_BiophysChem1999}.

Suppose we have an explicit solvent all-atom molecular system with
a total number of $N$ atoms, consisting of $N_{mol}$ solute atoms
with coordinates $\mathbf{r}$ (e.g., biomolecule) and $(N-N_{mol})$
solvent atoms with coordinates $\mathbf{w}$ (e.g., water atoms and
ions). Usually, an all-atom molecular mechanics force field, such
as AMBER~\citep{Amber20} or CHARMM,\citep{Brooks_JComputChem2009}
formulates a molecular potential function $v(\mathbf{r},\mathbf{w})$
as a sum of bonded and non-bonded terms.\citep{Adcock_ChemRev2006,Braun_LiveCoMS2018}
Therefore, without loss of generality, we can decompose $v(\mathbf{r},\mathbf{w})$
into three partial sums:\citep{Roux_BiophysChem1999} $v_{mol}(\mathbf{r})$
for interactions solely within and between the solute molecule(s),
$v_{w}(\mathbf{w})$ for those solely within and between solvent molecules
and $v_{mw}(\mathbf{r},\mathbf{w})$ for solute-solvent interactions:
\begin{equation}
v(\mathbf{r},\mathbf{w})=v_{mol}(\mathbf{r})+v_{w}(\mathbf{w})+v_{mw}(\mathbf{r},\mathbf{w}).\label{eq:Theory-decomp}
\end{equation}
We will refer to the solute-only potential $v_{mol}(\mathbf{r})$
as the ``vacuum potential'', since it only consists of terms that
describe the solute molecule(s) in vacuum.

For a chosen thermodynamic state (e.g., with fixed number of atoms
$N$, volume $V$ and temperature $T$ in a canonical ensemble), the
equilibrium probability density $p(\mathbf{r},\mathbf{w})$ for a
solute-solvent configuration $\mathbf{r},\mathbf{w}$ is:
\begin{equation}
p(\mathbf{r},\mathbf{w})=\frac{\mathrm{e}^{-\beta v(\mathbf{r},\mathbf{w})}}{\int\mathrm{d}\mathbf{r}\int\mathrm{d}\mathbf{w}\:\mathrm{e}^{-\beta v(\mathbf{r},\mathbf{w})}},\label{eq:Theory-p}
\end{equation}
where the scaling factor $\beta$ depends on the thermodynamic ensemble
used. In the canonical (NVT) ensemble at temperature $T$ it is given
by $\beta:=1/(k_{B}T)$ with Boltzmann constant $k_{B}$. The distribution
$p(\mathbf{r},\mathbf{w})$ can be sampled as a whole by MD or Monte
Carlo simulations with the explicit solvent potential $v(\mathbf{r},\mathbf{w})$.

For implicit solvent models, we are interested in recovering a potential
that describes the distribution of the solute molecules only. The
density associated with this potential is formed as the marginal density
obtained by integrating over the solvent degrees of freedom:
\begin{equation}
P(\mathbf{r}):=\int\mathrm{d}\mathbf{w}\:p(\mathbf{r},\mathbf{w}).\label{eq:Theory-P}
\end{equation}
We seek a potential function of solute coordinates $V(\mathbf{r})$
that could generate the marginal distribution $P(\mathbf{r})$. In
other words, the potential $V(\mathbf{r})$ should satisfy the following
equation:
\begin{equation}
\frac{\mathrm{e}^{-\beta V(\mathbf{r})}}{\int\mathrm{d}\mathbf{r}\:\mathrm{e}^{-\beta V(\mathbf{r})}}=P(\mathbf{r}).\label{eq:Theory-ThermoConsistency}
\end{equation}
By inserting Eq.~\eqref{eq:Theory-p} and~\eqref{eq:Theory-P} and
solving for $V(\mathbf{r})$, we have
\begin{equation}
V(\mathbf{r})=-\beta^{-1}\ln\left[\int\mathrm{d}\mathbf{w}\:\mathrm{e}^{-\beta v(\mathbf{r},\mathbf{w})}\right]+\textrm{const}.\label{eq:Theory-PMF}
\end{equation}
$V(\mathbf{r})$ is the so-called solute PMF,\citep{Roux_BiophysChem1999,Kirkwood_JChemPhys1935}
because its force corresponds to the mean force on the solute coordinates:
\begin{equation}
\mathbf{F}(\mathbf{r}):=-\nabla_{\mathbf{r}}V(\mathbf{r})=\left\langle \mathbf{f_{\mathbf{r}}}(\mathbf{r},\mathbf{w})\right\rangle _{\mathbf{r}},\label{eq:Theory-IS-meanforce}
\end{equation}
where
\begin{equation}
\mathbf{f}_{\mathbf{r}}(\mathbf{r,\mathbf{w}}):=\left[-\frac{\partial v}{\partial\mathbf{r}_{1}},\cdots,-\frac{\partial v}{\partial\mathbf{r}_{N}}\right]^{T}\label{eq:Theory-IS-force}
\end{equation}
denotes the forces on solute coordinates $\mathbf{r}$ with the solvent
conformation being $\mathbf{w}$, and
\[
\left\langle \cdot\right\rangle _{\mathbf{r}}:=\int\mathrm{d}\mathbf{w}\:\cdot p(\mathbf{r},\mathbf{w})
\]
is a marginal operator that averages over all solvent configurations
consistent with a given solute configuration according to the Boltzmann
distribution $p(\mathbf{r},\mathbf{w})$.

Theoretically, if we have $V(\mathbf{r})$ as defined in Eq.~\eqref{eq:Theory-PMF}
in the first place, then we can directly sample $P(\mathbf{r})$ as
in Eq.~\eqref{eq:Theory-P} and analyze most biologically relevant
processes, where solvent coordinates can be ignored (e.g., protein
folding, protein-ligand binding or in general any observable defined
by a function of the solute conformations only).\citep{Roux_BiophysChem1999}
However, in most cases we cannot solve the integral in Eq.~\eqref{eq:Theory-PMF}
analytically.

Alternatively, one can try to construct an approximation to the exact
PMF, which is usually determined by first fixing a range of candidates
with fixed functional forms $\{V(\mathbf{r};\Theta)\}$ and then optimizing
the parameter $\Theta$. An often adopted decomposition in the parameterization
is to separate the vacuum potential from the solvent-solvent and the
solute-solvent interactions. Applying Eq.~\eqref{eq:Theory-decomp}
to Eq.~\eqref{eq:Theory-PMF}, we can move $v_{mol}(\mathbf{r})$
out of the integral and thus
\begin{equation}
V(\mathbf{r})=v_{mol}(\mathbf{r})+V_{solv}(\mathbf{r}),\label{eq:Theory-IS-decomp}
\end{equation}
in which the \emph{solvation free energy} $V_{solv}$ is defined as
a function of solute configuration
\begin{equation}
V_{solv}(\mathbf{r}):=-\beta^{-1}\ln\left[\int\mathrm{d}\mathbf{w}\:\mathrm{e}^{-\beta\left[v_{w}(\mathbf{w})+v_{mw}(\mathbf{r},\mathbf{w})\right]}\right]+\textrm{const}.\label{eq:Theory-SFE}
\end{equation}
 Since the vacuum potential $v_{mol}(\mathbf{r})$ is known \textit{a
priori} from the all-atom force field, we can write any candidate
for approximating the solute PMF $V(\mathbf{r})$ in the following
form:
\begin{equation}
V(\mathbf{r};\Theta):=v_{mol}(\mathbf{r})+V_{solv}(\mathbf{r};\Theta),\label{eq:Theory-ISmodel}
\end{equation}
and optimizing $V(\mathbf{r};\Theta)$ is equivalent to finding the
best approximation $V_{solv}(\mathbf{r};\Theta^{*})$ to the solvation
free energy as defined in Eq.~\eqref{eq:Theory-SFE}. $V_{solv}(\mathbf{r};\Theta^{*})$
is an implicit solvent model, since it does not explicitly involve
any solvent, but can be used to approximately sample the Boltzmann
distribution of solute conformations by taking into account the solvent
environment implicitly according to Eq.~\eqref{eq:Theory-ThermoConsistency}.

\subsection{\label{subsec:Theory-TraditionalIS}Traditional implicit solvent
models}

A widely used strategy for parameterizing implicit solvent models
is to decompose the solvation free energy (Eq.~\eqref{eq:Theory-SFE})
into two terms: the non-polar $V_{solv}^{\textrm{np}}$ and the electrostatic~(polar)
$V_{solv}^{\textrm{elec}}$ contributions
\begin{equation}
V_{solv}(\mathbf{r})=V_{solv}^{\textrm{np}}(\mathbf{r})+V_{solv}^{\textrm{elec}}(\mathbf{r}),\label{eq:Theory-np-elec}
\end{equation}
and seek approximations for both terms separately (details can be
found in Ref.~\onlinecite{Roux_BiophysChem1999}). Various models
have been developed based on generalization of simple physical models
and/or heuristics~\citep{Chen_CurrOpinStructBiol2008,Feig_2008,Kleinjung_CurrOpinStructBiol2014}
for each of the two terms.

Here we illustrate Eq.~\eqref{eq:Theory-np-elec} through an example
of the popular generalized Born models.\citep{Chen_CurrOpinStructBiol2008,Feig_2008,Kleinjung_CurrOpinStructBiol2014}
As the name suggests, these models employ an approximation to the
electrostatics by generalizing the Born model~\citep{Born_ZPhys1920}
for charged spherical particles (e.g., simple ions):
\begin{subequations}
\begin{align}
V_{solv,\textrm{GB}}^{\textrm{elec}} & =\frac{1}{2}\left(\frac{1}{\epsilon_{out}}-\frac{1}{\epsilon_{in}}\right)\sum_{i,j}\frac{q_{i}q_{j}}{f_{ij}},\label{eq:Theory-GB-1}\\
\textrm{in which }f_{ij} & =\sqrt{r_{ij}^{2}+B_{i}B_{j}\exp\left(-\frac{r_{ij}}{4B_{i}B_{j}}\right)},\label{eq:Theory-GB-2}
\end{align}
\end{subequations}
 where $\epsilon_{out}$ and $\epsilon_{in}$ are outer and inner
(regarding the generalized Born sphere) dielectric constants, respectively.
Parameters $\left\{ q_{i}\right\} $, $\left\{ r_{ij}\right\} $ and
$\left\{ B_{i}\right\} $ denote the atomic partial charges, the pairwise
distances and the generalized Born radii, respectively.\citep{Onufriev_AnnuRevBiophys2019,Tsui_Biopolymers2000}
The non-polar contributions are typically represented by a linear
function of the solvent-accessible surface area~(SASA) is used to
represent the non-polar term
\begin{equation}
V_{solv,\textrm{SA}}^{\textrm{np}}=\gamma A(\mathbf{r})\;(+V_{0}^{\textrm{np}}),\label{eq:Theory-SA}
\end{equation}
in which $\gamma$ is a model parameter with the unit of surface tension
and $A(\mathbf{r})$ denotes the surface area associated with the
solute configuration $\mathbf{r}$ (sometimes a predetermined offset
$V_{0}^{\textrm{np}}$ is also used).\citep{Qiu_JPhysChemA1997,Levy_JAmChemSoc2003}
Generalized Born models together with a SASA-based non-polar treatment
form the so-called GBSA models, although other variants of non-polar
terms also exist.\citep{Roux_BiophysChem1999,Onufriev_AnnuRevBiophys2019}
Reference~\onlinecite{Onufriev_AnnuRevBiophys2019} provides a useful
review for the development and commonly used variants of generalized
Born models.

\subsection{\label{subsec:Theory-FM4IS}Implicit solvent model from a coarse-graining
point of view}

We put forward an alternative way for finding an approximation to
the solute PMF (Eq.~\eqref{eq:Theory-ThermoConsistency}) by adapting
the multi-scale coarse graining theory, which enables us to directly
optimize a candidate implicit solvent model against the conformations
and corresponding forces from explicit solvent simulations. Similar
ideas have been successfully applied to models of lipid bilayers~\citep{Izvekov_JPhysChemB2009}
and ionic solutions~\citep{Cao_JChemTheoryComput2013} under the
name of solvent-free coarse graining, but not to complex polymer systems,
such as peptide and proteins.

The multi-scale coarse graining theory was developed for parameterizing
potential functions for a CG system obtained through a linear CG mapping
that satisfies some general requirements (e.g., one atom cannot be
assigned to more than one CG bead).\citep{Izvekov_JPhysChemB2005,Noid_JChemPhys2008}
Since detailed derivations can be found in Ref.~\onlinecite{Noid_JChemPhys2008},
here we focus on its implications for the implicit solvation problem.

Consider a CG mapping $\Xi$ that treats each solute atom in a system
as a ``CG particle'':
\begin{equation}
\mathbf{r}=\Xi\mathbf{\left[\begin{array}{c}
\mathbf{r}\\
\mathbf{w}
\end{array}\right]},\;\Xi=\mathbf{\left[\begin{array}{c}
\mathbf{I}_{\mathrm{N_{mol}}}\\
\mathbf{0}
\end{array}\right]},\label{eq:Theory-CG-mapping}
\end{equation}
where this linear transformation essentially truncates the coordinates
by eliminating the solvent degrees of freedom. It is straightforward
to show that the CG system defined by the mapping $\Xi$ can be treated
under the multi-scale coarse graining framework, and the solute PMF
defined by Eq.~\eqref{eq:Theory-ThermoConsistency} is a CG PMF with
\emph{thermodynamic consistency}.\citep{Noid_JChemPhys2008} Moreover,
the mean force, $\mathbf{F}(\mathbf{r})$, acting on the solute (as
derived in Eq.~\eqref{eq:Theory-IS-meanforce}) is a CG mean force.

More than merely a change of notation, treating the implicit solvent
system as a CG system of the explicit one enables us to apply the
variational FM method for parameterizing an implicit solvent model.
For each candidate potential function $V(\mathbf{r};\Theta)$, the
multi-scale coarse graining functional~\citep{Noid_JChemPhys2008}
is defined as:
\begin{equation}
\chi\left[\Theta\right]:=\frac{1}{3N_{mol}}\left\langle \left\Vert \mathbf{f}_{\mathbf{r}}(\mathbf{r})+\nabla_{\mathbf{r}}V\left(\mathbf{r};\Theta\right)\right\Vert ^{2}\right\rangle ,\label{eq:Theory-MSCG-functional}
\end{equation}
where $\mathbf{f}_{\mathbf{r}}$ is defined in Eq.~\eqref{eq:Theory-IS-force},
$\left\Vert \cdot\right\Vert $ is the Frobenius norm and the bracket
$\left\langle \cdot\right\rangle $ indicates an average over a Boltzmann
distribution of fine grained configurations $(\mathbf{r},\mathbf{w})$.
The multi-scale coarse graining theory states that the global minimum
of this functional is unique (up to a constant) and corresponds to
the CG PMF $V(\mathbf{r})$, when the space of all possible functions
is considered.\citep{Noid_JChemPhys2008} Furthermore, within a given
family of functions parameterized as $\{V(\mathbf{r};\Theta)\}$,
one can variationally optimize the approximation by minimizing $\chi\left[\Theta\right]$.

Specifically for an implicit solvent model $V_{solv}(\mathbf{r};\Theta)$,
the multi-scale coarse graining functional can be rewritten into the
following form (implicit solvent functional) with the vacuum force
$\mathbf{f}_{mol}(\mathbf{r})=-\nabla_{\mathbf{r}}U_{mol}$:
\begin{equation}
\chi\left[\Theta\right]=\frac{1}{3N_{mol}}\left\langle \left\Vert \mathbf{f}_{\mathbf{r}}-\mathbf{f}_{mol}+\nabla_{\mathbf{r}}V_{solv}\left(\mathbf{r};\Theta\right)\right\Vert ^{2}\right\rangle .\label{eq:Theory-IS-functional}
\end{equation}

\subsection{\label{subsec:Theory-ML4IS}Machine learning of a CG model and of
an implicit solvent model}

Consider the parameterization of a CG force field:~we usually choose
a specific form of potential energy functions $\{V(\mathbf{r};\Theta)\}$
with trainable parameters $\Theta$, and then try to assign suitable
parameters $\Theta^{*}$, such that the model acquires desired accuracy
for representing the system of interest. The function form can be
either simple (e.g., Go-models) or complex (e.g., expressed by neural
networks). The performance assessment, i.e., the criteria for a good
model, can vary depending on the actual problem. However, after we
fix the functional form and the criterion (or a finite set of criteria)
to assess the ``suitability'' of a given model, the parameterization
procedure fits into the category of supervised machine learning. In
other words, we can approximate the CG PMF by numerically optimize
the trainable parameters $\Theta$.

In the context of multi-scale coarse graining, the criterion is the
functional defined in Eq.~\eqref{eq:Theory-MSCG-functional}. Although
its value usually cannot be directly computed analytically, we can
use a data-driven approximation in the minimization procedure:
\begin{equation}
\chi\left[\Theta\right]\approx L\left[\{\mathbf{r}_{i}\},\{\mathbf{f}_{i}\};\Theta\right]=\frac{1}{3NM}\sum_{i=1}^{M}\left\Vert \mathbf{f}_{i}+\nabla_{\mathbf{r}}V\left(\mathbf{r}_{i};\Theta\right)\right\Vert ^{2},\label{eq:Theory-MSCG-FMerror}
\end{equation}
which averages over a batch of CG coordinates $\{\mathbf{r}_{i}\}$
($M$ frames) and corresponding instantaneous forces $\{\mathbf{f}_{i}\}$
after CG mapping sampled from the thermodynamic equilibrium of the
fine-grained system. $L\left[\{\mathbf{r}_{i}\},\{\mathbf{f}_{i}\};\Theta\right]$
in Eq.~\eqref{eq:Theory-MSCG-FMerror} is often referred to as CG\textendash FM
error due to its mean-squared-difference form,\citep{Noid_JChemPhys2008,Noid_JChemPhys2008b}
and may serve as a loss function in the numerical optimization of
$\Theta$.

The CGnet method,\citep{Wang_ACSCentSci2019} for example, expresses
the candidate CG potential as an artificial neural network\footnote{In the regularized variant, there is also a prior energy to avoid
physically unfavorable states.\citep{Wang_ACSCentSci2019}} based on molecular features, such as pairwise distances, the angles
and dihedral angles formed by the CG particles. Since this potential
is fully determined by the neural network parameters, the optimization
of candidate function is equivalent to standard neural network training
in a supervised learning problem. Within this general framework, an
improved version\textemdash CGSchNet\textemdash has been developed
by using a more sophisticated graph neural network instead of multi-layer
perceptrons (see the next subsection for details).\citep{Husic_JChemPhys2020}

Similarly, from the implicit solvent functional (Eq.~\eqref{eq:Theory-IS-functional})
we can construct the implicit solvent FM loss function:
\begin{multline}
L\left[\{\mathbf{r}_{i}\},\{\mathbf{f}_{i}\};\Theta\right]=\\
\frac{1}{3NM}\sum_{i=1}^{M}\left\Vert \mathbf{f}_{i}-\mathbf{f}_{mol}\left(\mathbf{r}_{i}\right)+\nabla_{\mathbf{r}}V_{solv}\left(\mathbf{r}_{i};\Theta\right)\right\Vert ^{2},\label{eq:Theory-ISFMerror}
\end{multline}
where the $\{\mathbf{r}_{i}\}$ and $\{\mathbf{f}_{i}\}$ are coordinates
and forces for the solute from an equilibrated explicit solvent sample,
and $\mathbf{f}_{mol}$ for the vacuum force as defined in Eq.~\eqref{eq:Theory-IS-functional}.
An implicit solvent potential $V_{solv}\left(\mathbf{r}_{i};\Theta\right)$
can thus be learned for a given molecular system using a given optimizable
model (e.g., a neural network).

\subsection{\label{subsec:ISSNet-architecture}The ISSNet architecture}

\begin{figure*}
\centering{}\includegraphics[width=1\textwidth]{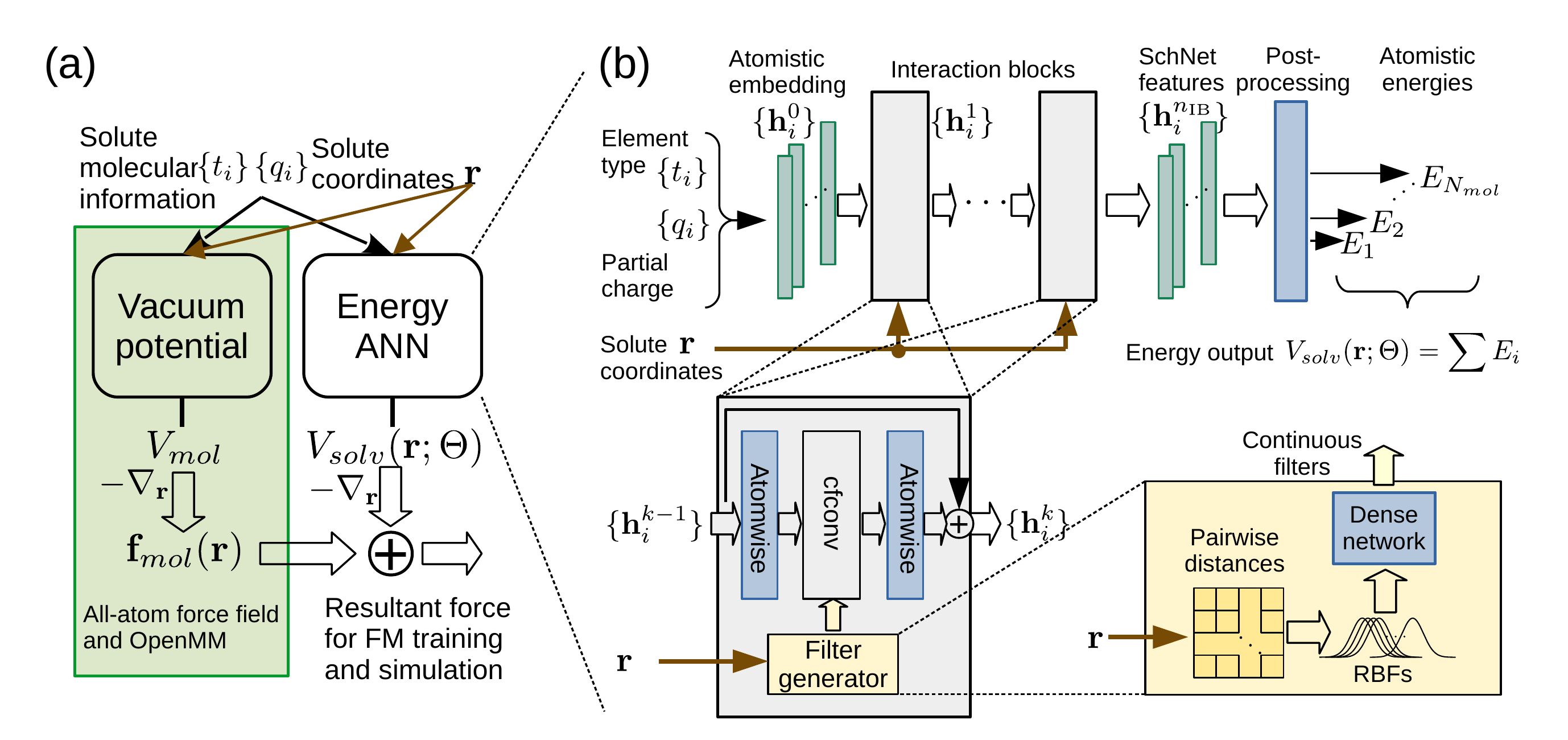}\caption{\label{fig:ISSchNet}Schematic representations of the ISSNet: (a)
overall architecture, (b) the detailed structure of neural network.}
\end{figure*}

We construct a specific artificial neural network architecture for
the deep learning of an implicit solvent model \textemdash ISSNet
(a shorthand for \uline{i}mplicit \uline{s}olvent \uline{S}ch\uline{Net}).
Fig.~\ref{fig:ISSchNet}a illustrates the architecture of the ISSNet
with the left and right columns corresponding to the vacuum potential
$v_{mol}(\mathbf{r})$ and the solvation energy $V_{solv}(\mathbf{r};\Theta^{*})$
as in Eq.~\eqref{eq:Theory-ISmodel}, respectively. The vacuum potential
and forces in the lime-colored box come directly from the all-atom
force field, and are thus irrelevant to the training process. On the
right side, the core of ISSNet is an energy network, which can be
regarded as a function that receives all-atom 3D coordinates $\mathbf{r}$
of the solute molecule(s) and returns a single energy scalar $V_{solv}(\mathbf{r};\Theta)$.
The functional relation between $V_{solv}$ and $\mathbf{r}$ is determined
by the neural network and its trainable parameters $\Theta$. When
the functional relation meets certain smooth requirements, it immediately
provides a force field $F=-\nabla_{\mathbf{r}}V_{solv}(\mathbf{r};\Theta)$
for MD simulation.

We follow the CGSchNet architecture~\citep{Husic_JChemPhys2020}
and employ SchNet~\citep{Schutt_JChemPhys2018} to express $V_{solv}$.
SchNet is a type of graph neural network for molecular systems.\citep{Schutt_JChemPhys2018}
It maps each atom (or CG particle in CGSchNet~\citep{Husic_JChemPhys2020})
to a node in a graph, and we can subsequently define edges for node
pairs based on the proximity in the 3D space. When we use a uniform
distance cutoff and uses a shared sub-neural network to generate the
edge information, the graph representation will enable the SchNet
to learn of molecular representations while enforcing the translational
and rotational symmetries of molecular potentials. Furthermore as
stated in Ref.~\onlinecite{Husic_JChemPhys2020}, it lays a foundation
for model transferability across different molecular systems (see
also the Discussion section).

Figure~\ref{fig:ISSchNet}b shows the data flow in a SchNet:\citep{Schutt_JChemPhys2018}
a starting feature vector (i.e., the \emph{embedding}) $\mathbf{h}_{i}^{0}$
is generated for each node. Each \emph{interaction block} updates
the atomistic feature $\left\{ \mathbf{h}_{i}^{k}\right\} $ to $\left\{ \mathbf{h}_{i}^{k+1}\right\} $
by summarizing the information on the neighboring nodes through continous-filter
convolution (cfconv). By stacking multiple ($N_{\textrm{IB}}$) interaction
blocks, information can be propagated farther among the nodes to express
longer-ranged and/or sophisticated interactions. Afterwards, a \emph{post-processing
sub-network} maps the feature $\left\{ \mathbf{h}_{i}^{N_{\textrm{IB}}}\right\} $
on each atom/bead into a scalar atomistic energy. Finally, the energy
contributions from each atom are summed up to produce the total energy
prediction, which in our case is used to express the implicit solvent
potential $V_{solv}\left(\mathbf{r}_{i};\Theta\right)$.

The generation of embedding vectors for the system is an important
step to incorporate useful chemical and physical information that
we know \textit{a priori} for each atom. In this work we use three
variants of ISSNet for parameterizing an implicit solvent potential
(shown in Fig.~\ref{fig:ISSchNet}b):
\begin{enumerate}
\item The first variant (denoted as ``t-ISSNet'') follows the original
SchNet scheme, i.e., distinguishing the atoms by their nuclear charges.\citep{Schutt_JChemPhys2018}
In this case, only the information about element types $\left\{ t_{i}\right\} $
is used. This vector comprises the nuclear charge for each solute
atom, thus using a unique natural number to denote each element. The
embedding for the $i$-th atom $\mathbf{h}_{i}^{0}$ is taken from
the $t_{i}$-th row of a trainable matrix $A$: $\mathbf{h}_{i}^{0}=A_{t_{i}}$.
\item The second (``q-ISSNet'') is inspired by the generalized Born models,
which entail not only a parameter specified by the atom type, but
also include the atomic partial charge from the force field in the
potential expression. In practice, we encode the partial charge (divided
by the elementary charge unit) $q_{i}$ of each atom into a vector:
\[
\mathbf{e}(q_{i})=\textrm{Dense-Net}\left(\textrm{RBF}\left(q_{i};\mathbf{\mu},\gamma\right)\right),
\]
in which the $\textrm{Dense-Net}$ is a dense neural network, the
radial basis function (RBF) vector is defined as:
\begin{equation}
\textrm{RBF}\left(q;\mathbf{\mu},\gamma\right)=\left[e^{-\gamma(q-\mu_{k})^{2}}\right]^{\top},\label{eq:Methods-RBF}
\end{equation}
with the entries in $\mathbf{\mu}\in\mathbb{R}^{N_{c}}$ uniformly
placed over the range $[-1,1]$ (covering all possible partial charge
value for atoms in amino acids) and $N_{\textrm{c}}$, $\gamma$ are
hyperparameters. Based on this newly introduced embedding function
$\mathbf{e}(\cdot)$ and partial charge information $\left\{ q_{i}\right\} $,
we use charge embedding $\mathbf{e}(q_{i})$ instead of the atomic-type
embedding as in ``t-ISSNet'' as the initial feature.
\item The third (``qt-ISSNet'') is a mixture of the above variants. Both
the type and charge embeddings are calculated and then concatenated
into a mixed feature vector for each atom. Note that the sub-vectors
$A_{t_{i}}$ and $\mathbf{e}(q_{i})$ have only half of the normal
length of the above two embeddings, such that the output vector still
keep the same width.
\end{enumerate}
Once the embedding is generated, each atom receives a starting feature
vector. The interaction blocks then perform continuous-filter convolutions
(cfconv) over the feature vectors.\citep{Schutt_JChemPhys2018} The
distance between each neighboring node pair $i$ and $j$ is expanded
in a RBF vector (defined in Eq.~\eqref{eq:Methods-RBF}), which is
in turn featurized into a ``continuous filter'' by a dense network:
\begin{equation}
\mathbf{e}_{ij}=\textrm{\textrm{Dense-Net}\ensuremath{\left(\textrm{RBF}\left(\left|\mathbf{r}_{i}-\mathbf{r}_{j}\right|;\mathbf{\mu}_{d},\gamma_{d}\right)\right)}},
\end{equation}
where $\gamma_{d}$ and $\mathbf{\mu}_{d}\in\mathbb{R}^{N_{\textrm{RBF}}}$
are pre-selected hyperparameters. For each node $i$, the cfconv is
performed upon the feature vectors:
\begin{equation}
\mathbf{y}_{j}^{l}\mapsto\sum_{j}\mathbf{e}_{ij}\odot\mathbf{y}_{j}^{l},
\end{equation}
where $\odot$ denotes elementwise multiplication. In addition, dense
networks (also known as atomwise layers in Ref.~\onlinecite{Schutt_JChemPhys2018}
and in Fig.~\ref{fig:ISSchNet}b) with trainable weights and biases
act on the feature vectors before and after the cfconv operation,
which gives additional functional expressivity to the transformation
of feature vectors. To avoid vanishing gradients, the output of the
$l$-th interaction block is summed with the input $\left\{ \mathbf{h}_{i}^{l}\right\} $
following a residual network scheme. Putting them all together, the
update in the $l$-th interaction block can be expressed as:
\begin{equation}
\mathbf{h}_{i}^{l+1}=\mathbf{h}_{i}^{l}+\textrm{AW}_{\textrm{post}}^{l}\left[\sum_{j}\mathbf{e}_{ij}\odot\textrm{AW}_{\textrm{pre}}^{l}\left(\mathbf{h}_{j}^{l}\right)\right],
\end{equation}
where the $\textrm{AW}$s are atomwise layers.

Apart from the variants of embedding generations, there are other
hyperparameters for a ISSNet model. Examples include the width of
the feature vectors $W$, the number of interaction blocks $N_{\textrm{IB}}$,
the number and distribution of RBF centers $\vec{\mu}_{d}$. Hyperparameters
have to be fixed before training a certain model, but the choice can
be optimized through cross validation.

\subsection{\label{subsec:TrainValSimu}Training, validation of and simulation
with an ISSNet model}

Given an ISSNet and the implicit solvent FM loss function Eq.~\eqref{eq:Theory-ISFMerror},
we follow the typical training procedure for a supervised deep learning
problem,\citep{LeCun_Nature2015,Hastie_2009} which is also used for
CGnet~\citep{Wang_ACSCentSci2019} and CGSchNet:\citep{Husic_JChemPhys2020}
\begin{enumerate}
\item Separate the available data (recorded in equilibrium sampling of an
explicit solvent system) into training and validation sets.
\item Repeat for a fixed number of epochs:
\begin{enumerate}
\item Randomly shuffle the solute coordinates and corresponding forces $\{(\mathbf{r}_{i},\mathbf{f}_{i})\}$
for training.
\item Split the training data into small batches with a pre-determined size
$M$.
\item For each batch:
\begin{enumerate}
\item Evaluate the FM loss $L\left[\{\mathbf{r}_{i}\},\{\mathbf{f}_{i}\};\Theta\right]$
on the batch.
\item Update the model parameters $\Theta$ by applying a stochastic gradient
descent method (e.g., the Adam optimizer~\citep{Kingma_arXiv2014}).
\end{enumerate}
\item Evaluate the FM loss on the validation set.
\end{enumerate}
\end{enumerate}
We choose suitable hyperparameters for our models based on cross validation:
we divide the data set into four equal parts after shuffling. Then
we conduct four rounds of independent model training with the same
setup, each round with a different fold serving as validation set
and the other three as training set. The cross validation force matching
(CV\textendash FM) error is calculated by averaging validation errors
from the four training processes, which is considered as a reliable
benchmark of the chosen hyperparameter set.\citep{Wang_ACSCentSci2019,Husic_JChemPhys2020}
For example in Ref.~\onlinecite{Wang_ACSCentSci2019}, it was shown
that this error corresponded well to the free energy difference metrics
after sampling with trained CG models. Therefore, we performed hyperparameter
searches by comparing the CV\textendash FM errors among a series of
hyperparameters (see SI, Section~B).

Trained ISSNet models can be used for implicit solvent simulations.
We perform such simulations with the MD simulation library OpenMM~\citep{Eastman_PLoSComputBiol2017}
and a plugin for incorporating a PyTorch model as force field:\citep{openmm_torch}
evaluate the forces from both the neural network $V_{solv}(\mathbf{r};\Theta^{*})$
and the vacuum potentials $V_{mol}$ at each time step, and then perform
simulation with the resultant force on the solute molecule. Section~A
of the SI describes the simulation setup, which resembles that of
the explicit solvent simulation for the generation of training data
sets. For a review of the basic MD concepts and conventions, we refer
the readers to comprehensive reviews, such as Refs.~\onlinecite{Adcock_ChemRev2006}
and.~\onlinecite{Braun_LiveCoMS2018}.

For an accurate evaluation the thermodynamics of implicit solvent
systems, we need to sample sufficiently many conformations according
to the Boltzmann distribution. In this study we achieve this by aggregating
multiple long MD trajectories. We leverage batch-evaluation of neural
network forces by simulating with several replicas of the same system
in parallel, which significantly reduces the time needed to achieve
a long cumulative simulation time for our test molecular systems.
Similar strategies have been used to obtain the converged thermodynamics
of coarse grained systems with CGnet/CGSchNet.\citep{Husic_JChemPhys2020,Wang_ACSCentSci2019}
We also incorporate PT\textendash MD~\citep{Swendsen_PhysRevLett1986,Sugita_ChemPhysLett1999}
as an enhanced sampling method~\citep{Zuckerman_AnnuRevBiophys2011,Bernardi_2015}
so as to assist transitions among metastable states for the chignolin
system. Implementation of a general-purpose tool for batch simulations
with optional PT exchanges can be found in Ref.~\footnote{Code that facilitated the experiments in the paper: \texttt{\href{https://github.com/noegroup/reform/releases/tag/v0.1}{https://github.com/noegroup/reform/releases/tag/v0.1}.}}.

\section{Results}

To assess the usability and performance of our neural-network-based
implicit solvent method, we train models for two molecular systems\textemdash capped
alanine and chignolin\textemdash and use the trained models in implicit
solvent simulations. These two systems were also used as examples
and benchmarks for CGnet and CGSchNet.\citep{Husic_JChemPhys2020,Wang_ACSCentSci2019}
We then compare the free energy landscapes implied by the output trajectory
from the reference all-atom simulation, implicit solvent simulations
with our model, and those with a widely used GBSA model.\citep{Onufriev_Proteins2004}
The comparison shows that our model outperforms the classical model
in terms of recovering the thermodynamics of explicitly solvated systems.

\subsection{Capped alanine\label{subsec:Results-ala2}}

Capped alanine, also known as alanine ``dipeptide'', has two essential
degrees of freedom: the torsion angles $\phi$ (C\textminus N\textminus C$\alpha$\textminus C)
and $\psi$ (N\textminus C$\alpha$\textminus C\textminus N).\citep{Apostolakis_JChemPhys1999,Feig_JCTC2008,Pettitt_ChemPhysLett1985,Tobias_JPhysChem1992}
Consisting of only 22 atoms, it is a simple yet meaningful system
in many studies, e.g., conformational analyses,\citep{Anderson_Proteins1988,Apostolakis_JChemPhys1999,Nuske_JChemPhys2019}
free energy surface calculations~\citep{Apostolakis_JChemPhys1999,Feig_JCTC2008,Pettitt_ChemPhysLett1985,Tobias_JPhysChem1992}
and solvation effects.\citep{Drozdov_JACS2004,Pettitt_ChemPhysLett1985,Tobias_JPhysChem1992}
Here we expect a good implicit solvent model to reproduce the conformational
density distribution in a simulation of capped alanine on the $\phi-\psi$
plane (i.e., a Ramachandran map) as given by the explicit solvent
simulations.

To prepare a data set for model training and validation, we performed
a 1-$\mu$s all-atom molecular dynamics simulation of a capped alanine
molecule with TIP3P explicit solvent model (see SI, Section~A). The
conformations and corresponding instantaneous all-atom forces on the
solute (capped alanine) atoms were collected every picosecond to form
the data set, forming a data set with $10^{6}$ samples. We randomly
shuffle the collected coordinate-force pairs and divide them to four
folds of equal sizes.

We train and validate ISSNet implicit solvent models for capped alanine
on the prepared data set with the FM scheme introduced in the Theory
and methods section. The training and validation process (detailed
setup in the SI, Section~B) of our ISSNet solvent models are comparable
to those of a standard CGnet~\citep{Wang_ACSCentSci2019} or CGSchNet.\citep{Husic_JChemPhys2020}
Essentially, we set aside one fold of the available data for validation,
and mix the data from the rest three folds for training. We also performed
four-fold cross validations for sets of hyperparameters (listed in
Table~S2) to observe how they affect the learning and prediction
of the solvation mean force. By comparing the mean CV\textendash FM
errors for each condition (see Fig.~S1 in the SI), we conclude that
the force prediction accuracy of trained models is in general robust
to most hyperparameter settings (comparable to the findings in Ref.~\onlinecite{Husic_JChemPhys2020}).
The only hyperparameter that significantly influenced the CV\textendash FM
error was the embedding: type-only (t-), charge-only (q-) or type-and-charge
(qt-), among which the partial charge-only variant (q-ISSNet) produced
the lowest CV\textendash FM error. We selected the model with the
lowest validation FM error for each embedding setup in the cross-validation
processes for further analyses.

We perform simulations of the capped alanine system with each trained
implicit solvent model to examine its performance. In order to accumulate
enough samples in the conformational space in a relatively short time,
we performed simulations in batch mode starting from 96 conformations.
The starting structures were sampled from the all-atom simulation
trajectory based on the equilibrium distribution, which was in turn
estimated by a Markov State Model (MSM)~\citep{Prinz_JChemPhys2011}
with the PyEMMA software package.\citep{wehmeyer2018introduction,scherer_pyemma_2015}
The full setup for implicit solvent simulations can be found in the
SI, Section~A. In addition, we ran implicit solvent simulation with
a traditional GBSA model for comparison. We used the default model
(GBSA\textendash OBC) provided by the OpenMM suite~\citep{Eastman_PLoSComputBiol2017}
for AMBER force fields,\citep{Amber20} which is based on the work
of Onufriev et al.~\citep{Onufriev_Proteins2004} The same set of
Boltzmann-distributed starting structures was used in batch simulations
to ensure the comparability of the results across different solvent
models.

\begin{figure*}
\centering{}\includegraphics{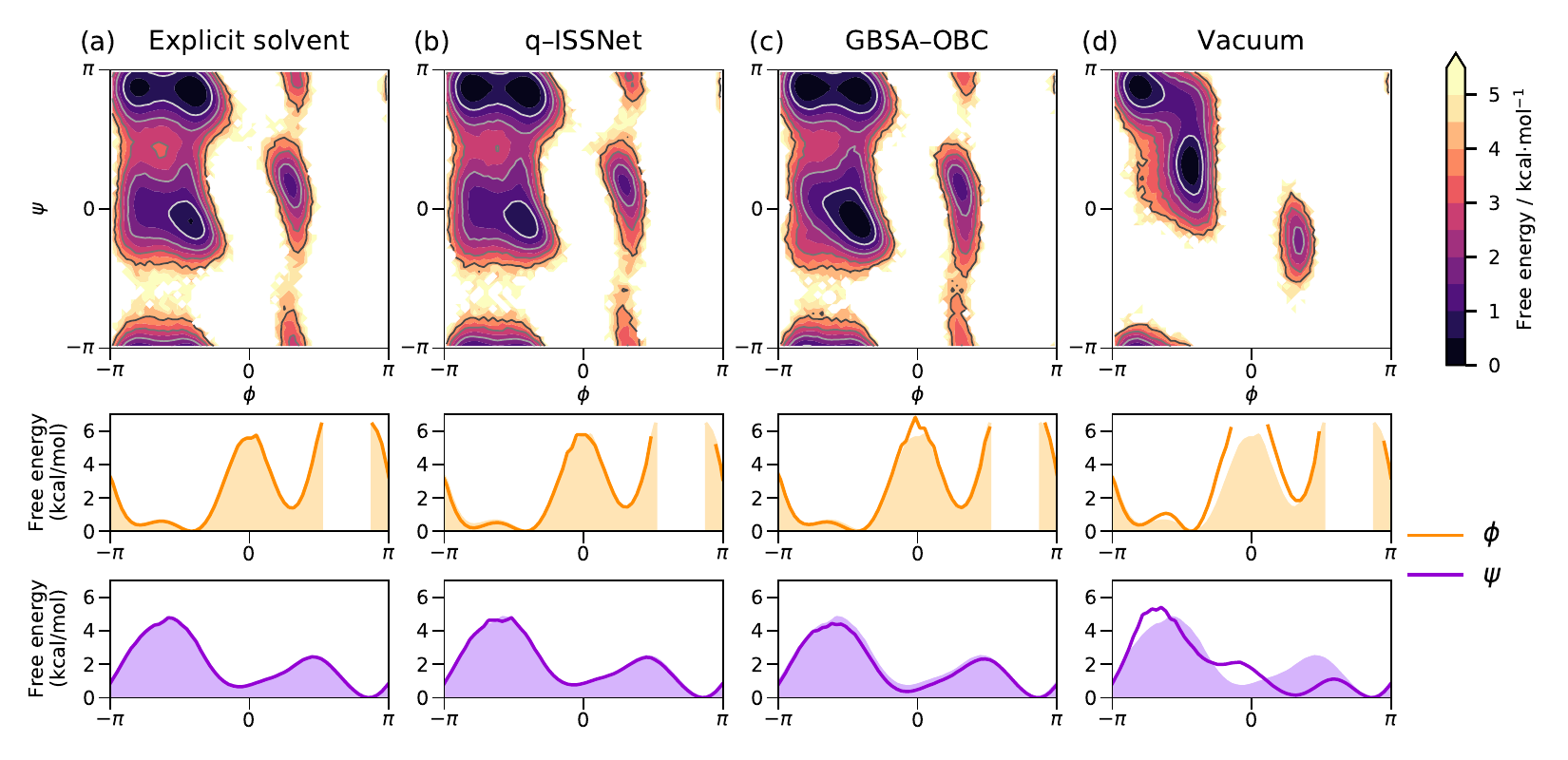}\caption{\label{fig:Results-ala2}Two- and one-dimensional free energy plots
for all-atom capped alanine systems: (a) explicit solvent system with
TIP3P water model (reference), the implicit solvent setup with (b)
trained q\textendash ISSNet and (c) the GBSA\textendash OBC model,
and (d) the vacuum system without solvation treatment (used as a negative
control). The 2D free energy surfaces are created by histogramming
of simulation trajectories on $\phi$- and $\psi$-dihedral angles
with MSM-reweighting, while the two 1-d free energy curves (bold lines)
below each contour plot show the corresponding marginal distributions.
For clear comparison of the 1-d distributions between the reference
system and the rest, we let the shaded regions represent the explicit
solvent result.}
\end{figure*}
By comparing the free energy landscapes with those from the reference
explicitly solvated and vacuum systems, we can assess how well the
implicit solvent model can approximate the solvent effects on thermodynamics.
Figure~\ref{fig:Results-ala2} shows the free energy surfaces for
the implicit solvent, reference explicit solvent and vacuum systems.
The free energy plots for systems with trained t-ISSNet and qt-ISSNet
models can be found in Section S2 in the SI. Qualitatively, the free
energy landscapes of the implicit solvent simulations (Fig.~\ref{fig:Results-ala2}~b
and \ref{fig:Results-ala2}~c) are dramatically different from the
vacuum case (Fig.~\ref{fig:Results-ala2}d), and recovers the main
energy minima emerging in the explicit solvent simulation (Fig.~\ref{fig:Results-ala2}a).
The sample proportion in these regions in implicit solvent simulations
also appears similar to the distribution for the explicit solvent
system, both on the 2D free energy landscapes and on the marginal
distributions for $\phi$ and $\psi$. The result proves that either
of the two implicit solvent models can properly model the solvent
effect, which is absent in the vacuum simulation. Between the implicit
solvent systems (with our trained neural network model and with the
GBSA\textendash OBC model), it is observed that the q-ISSNet model
corresponds to a free energy contour that better resembles the explicit
solvent reference. The other two variant of ISSNet models, although
gives slightly less accurate free energy lanscapes, but still outperform
the GBSA\textendash OBC model (see Fig.~S2 in the SI).

\begin{table}
\caption{\label{tab:Results-ala2-metrics}KL divergence, JS divergence and
MSE of free energy for comparing the discrete conformational distributions
on the $\phi-\psi$ plane of the implicit solvent, vacuum and the
explicit solvent systems for capped alanine. Calculation is performed
over the simulation trajectories after MSM reweighting (details in
the SI, Section~C). Bold font is designated for the lowest divergence/error
values, which correspond to the implicit solvent model with ISSNet
plus partial charge-only (q-) embeddings.}

\begin{ruledtabular}
\begin{tabular}{lr@{\extracolsep{0pt}.}lr@{\extracolsep{0pt}.}lr@{\extracolsep{0pt}.}l}
System & \multicolumn{2}{c}{$D_{\textrm{KL}}$\footnote{{\small{}Calculated in exactly the same manner as in Ref.}~\onlinecite{Husic_JChemPhys2020}{\small{}
and thus comparable to the KL divergence values reported there.}}$/10^{-2}$} & \multicolumn{2}{c}{\multirow{1}{*}{$D_{\textrm{JS}}$$/10^{-3}$}} & \multicolumn{2}{c}{$\mathrm{MSE}$\footnote{Unit:~(kcal/mol)$^{2}$;{\small{} calculation is done in the same
manner as in Ref.}~\onlinecite{Husic_JChemPhys2020}.}$/10^{-2}$}\tabularnewline
\colrule Explicit solvent & (0&) & (0&) & (0&)\tabularnewline
 & \multicolumn{2}{c}{} & \multicolumn{2}{c}{} & \multicolumn{2}{c}{}\tabularnewline
t\textendash ISSNet & 2&32 & 5&62 & 8&28\tabularnewline
\textbf{q\textendash ISSNet}\footnote{Used for comparison with reference systems in Fig.~\ref{fig:Results-ala2}.} & \textbf{1}&\textbf{46} & \textbf{3}&\textbf{63} & \textbf{7}&\textbf{64}\tabularnewline
qt\textendash ISSNet & 5&61 & 13&4 & 9&44\tabularnewline
 & \multicolumn{2}{c}{} & \multicolumn{2}{c}{} & \multicolumn{2}{c}{}\tabularnewline
GBSA\textendash OBC & 9&47 & 23&4 & 19&2\tabularnewline
 & \multicolumn{2}{c}{} & \multicolumn{2}{c}{} & \multicolumn{2}{c}{}\tabularnewline
Vacuum & 169& & 530& & 250&\tabularnewline
\end{tabular}
\end{ruledtabular}

\end{table}

The difference between implicit solvent models can be better analyzed
by directly comparing the discretized equilibrium distributions (i.e.,
the histograms) on the dihedral plane, which we used to generate the
free energy contours above. We evaluate the Kullback-Leibler (KL)
and Jensen-Shannon (JS) divergences between the distributions of various
models and that of the reference distribution, as well as the mean
squared error (MSE) of discrete free energies. Table~\ref{tab:Results-ala2-metrics}
presents these quantitative metrics for measuring the similarity of
the free energy landscapes between the implicit solvent or the baseline
vacuum system with the reference explicit solvent case. All three
columns give the same trend:~ISSNet implicit solvent models have
the smallest, vacuum energies the largest errors, with the GBSA\textendash OBC
implicit solvent model in between. This is consistent with the visual
comparison of the free energy surfaces in Fig.~\ref{fig:Results-ala2},
and indicates that our machine-learned implicit solvent method outperforms
the traditional GBSA\textendash OBC model for this system. Additionally,
the q-ISSNet variant (with charge-only embedding) corresponds to the
smallest difference from the reference among the ISSNet models.

\subsection{Chignolin\label{subsec:Results-Chignolin}}

Due to their small size and short folding time, the artificially designed
miniprotein chignolin~\citep{Honda_Structure2004} and its stabler
variant CLN025,\citep{Honda_JACS2008} are widely used as example
systems in both experimental~\citep{Davis_JACS2012,Honda_JACS2008}
and computational investigations~\citep{Satoh_FEBSLett2006,Suenaga_ChemAsianJ2007,Lindorff_Science2011,Nguyen_JACS2014}
of protein folding and kinetics. Additionally, thanks to the availability
of extensive reference data from experiments,\citep{Honda_JACS2008}
chignolin variants serve as benchmark systems in the development of
all-atom force fields~\citep{Maier_JCTC2015ff14sb,Harder_JCTC2016opls3,Huang_NatMethods2017charmm36m}
and for comparison among force fields~\citep{Kuhrova_BiophysJ2012}
and solvent methods.\citep{Anandakrishnan_BiophysJ2015} In this section
we use the CLN025 variant~\citep{Honda_JACS2008} of chignolin, a
10-amino-acid miniprotein with sequence YYDPETGTWY (together with
N- and C-terminal caps) as the solute molecule, which is referred
to simply as chignolin in the text below. The explicit solvent all-atom
simulation trajectories (available online~\citep{Perez_FigShare2021})
and corresponding force data were kindly provided by the authors of
Ref.~\onlinecite{Wang_ACSCentSci2019}. The simulation setup is reported
in the SI, Section~A. We randomly selected $2\times10^{5}$ coordinate\textendash solvation
force pairs from the aggregated data set (with $1.8\times10^{6}$
pairs in total) according to the equilibrium conformational distribution
estimated by a MSM for training and validation of ISSNet models, and
divide them to four folds of equal sizes.

The training and cross-validation procedures for chignolin are similar
to those for capped alanine with slightly modified setups (see Section~B
of the SI). In addition to the embedding choices, the number of interaction
blocks also appears to be influential to the CV\textendash FM errors
in hyperparameter searches (see Table~S3). Therefore, we trained
the ISSNet models with the three different types of embeddings and
two or three interaction blocks, resulting in six implicit solvent
models for the next step.

We performed vacuum and implicit solvent simulations for chignolin
(the latter with the trained ISSNet models), similar to those for
capped alanine. In order to facilitate transitions among metastable
states and thus a more accurate estimate of the state population with
multiple short-time simulations, we applied parallel tempering (PT)
methods in the MD simulations. We also performed a simulation with
the GBSA\textendash OBC model, and compared the outcome with those
corresponding to the ISSNet models. All simulations were initiated
from the same 16 starting structures, while were sampled from the
with MSM weights. More information regarding the simulation setups
can be found in the SI, Section~A.

\begin{figure*}
\centering{}\includegraphics{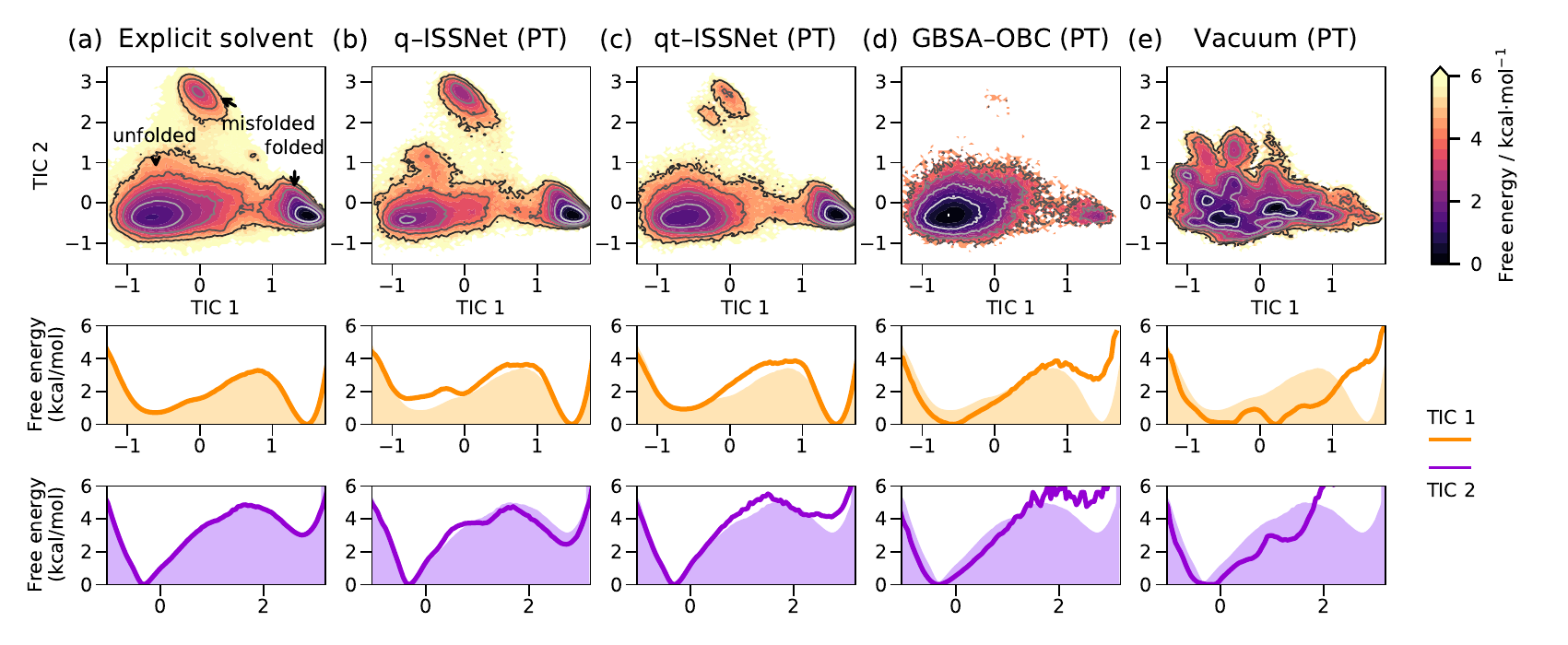}\caption{\label{fig:Results-chi}Two- and one-dimensional free energy plots
for all-atom chignolin systems: (a) explicit solvent system with mTIP3P
water model (reference), the implicit solvent setup with (b) trained
q\textendash ISSNet, (c) trained qt\textendash ISSNet and (d) the
GBSA\textendash OBC model, and (e) the vacuum system without solvation
treatment (negative control group). The 2D free energy surfaces are
created by histogramming of simulation trajectories on the first and
second TICs after TICA transformation. For the explicit solvent data
set, a MSM is estimated upon the short simulation trajectories, and
then used for reweighting in the histogram. For simulation with ISSNet
models or the vacuum simulation, we use PT\textendash MD to increase
state-transition rates. The two 1-d free energy curves (bold lines)
below each contour plot show the corresponding marginal distributions.
For clear comparison of the 1-d distributions between the reference
system and the rest, the shaded regions represent the explicit solvent
result from column a.}
\end{figure*}

In order to visualize the conformational distribution, we performed
time-lagged independent component analysis (TICA)~\citep{Perez_JChemPhys2013,Schwantes_JChemTheoryComput2013}
on the explicit solvent trajectories according to Ref.~\onlinecite{Husic_JChemPhys2020}
(over the pairwise C$_{\alpha}$ distances),\citep{Husic_JChemPhys2020}
and used the resulting TICA matrix to project the simulation results
for each model onto the same set of collective coordinates. The first
two time-lagged independent components (TICs) resolve the three metastable
states (see Fig.~\ref{fig:Results-chi}; cf.~figures in Ref.~\onlinecite{Husic_JChemPhys2020}).
Furthermore, a MSM is estimated on the explicit solvent simulation
data to obtain the correct weights for each frame in the trajectories,
such that we can more precisely estimate the free energy landscape
at equilibrium by histogramming (also used for capped alanine). Free
energy estimates for other systems in the comparison does not require
MSM-reweighting, since a sufficient and correct sampling from the
Boltzmann distribution is obtained by means of the PT simulation.
Apart from the change of coordinates, the plotting procedure (see
Section~C of the SI) is the same as described for capped alanine.

\begin{figure*}
\centering{}\includegraphics{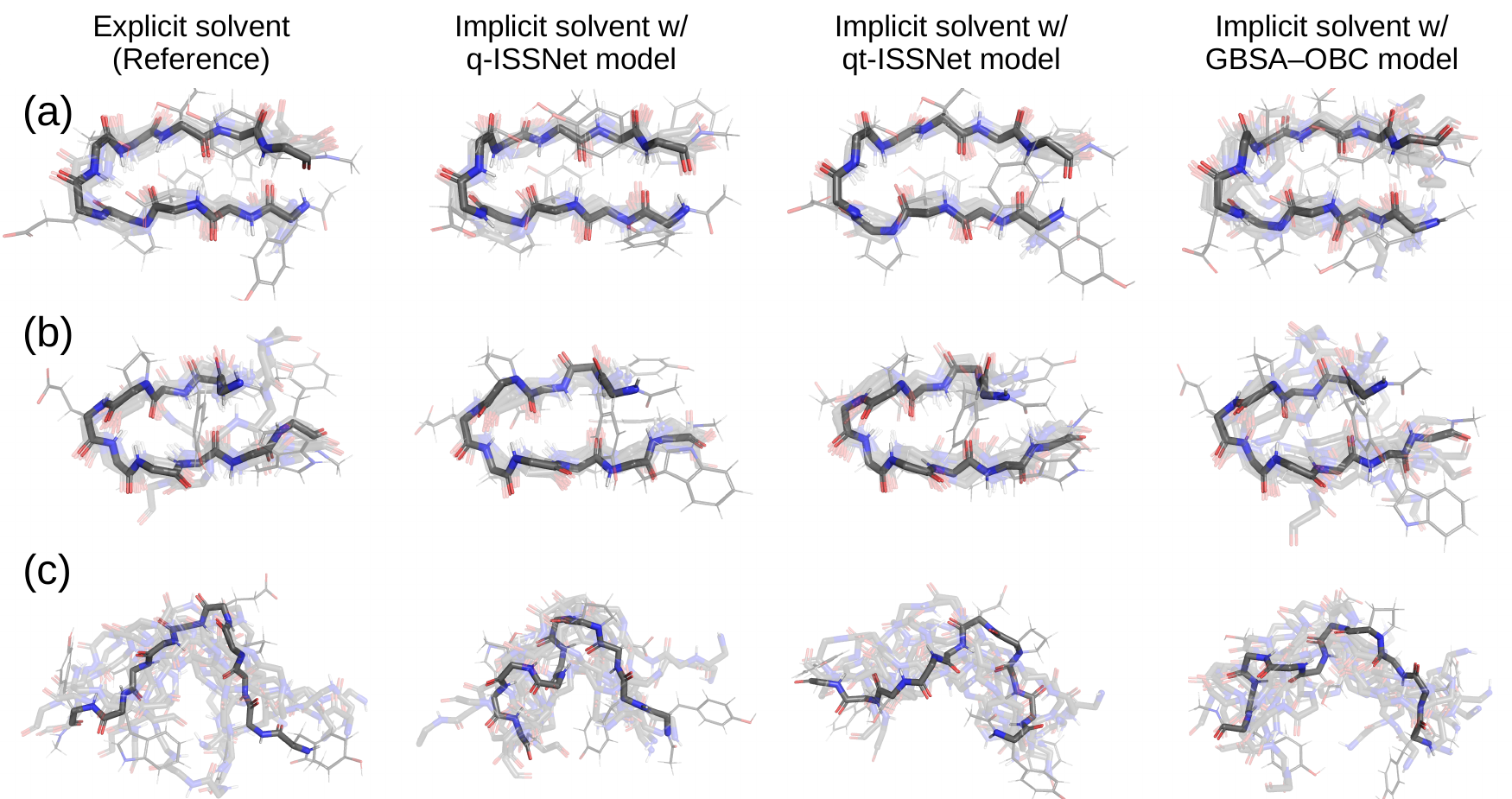}\caption{\label{fig:chiStructures}Representative structures of chignolin from
explicit and implicit solvent simulations from (a) the folded, (b)
the misfolded and (c) the unfolded metastable states. We overlay 10
structures randomly sampled from each metastable state for each solvent
model (cf.~Fig.~\ref{fig:Results-chi}) and visualize their backbone
structures. We highlight one structure in each plot and plot its side
chains in addition.}
\end{figure*}

Figure~\ref{fig:Results-chi} displays the equilibrium free energy
landscapes for two ISSNet models and the reference systems we introduced
above. For the convenience of description, we label the three major
minima on the free energy landscape in Fig.~\ref{fig:Results-chi}a
as ``misfolded'' (upper), ``unfolded'' (lower left) and ``folded''
(lower right) according to the folding status of the peptide conformations
in these minima. These minima correspond to metastable states from
MSM analyses~\citep{Husic_JChemPhys2020} (for details see the SI,
Section~D). By comparing the 2D free energy plots in Fig.~\ref{fig:Results-chi},
we can qualitatively conclude that the three metastable states are
present at the correct positions for all presented implicit solvent
systems (Fig.~\ref{fig:Results-chi}b\textendash d), although the
misfolded state is rarely visited in the GBSA\textendash OBC system.
Meanwhile, the vacuum system has an extremely rugged free energy landscape
mostly located in the unfolded region (Fig.~\ref{fig:Results-chi}e).
This shows that the implicit solvent models incorporate non-trival
solvent effects that are absent from the vacuum system. Another observation
is that the ISSNet models better reproduce the populations of the
folded and misfolded states, which are underestimated by the GBSA\textendash OBC
model. As a side note, a similar deficiency in the folded state population
for chignolin has been reported and analyzed for simulation with an
AMBER force field~\citealp{Amber20} and the GBSA\textendash OBC
model.\citep{Anandakrishnan_JChemTheoryComput2018}

In Fig.~\ref{fig:chiStructures} we visualize some representative
3D structures of chignolin sampled from the simulation trajectories.
We randomly pick 10 structures that were assigned to different metastable
states on the TIC1\textendash TIC2 plot (for details see the SI, section
S4) and only plot the backbone atoms for clarity. We randomly pick
one from the 10 structures for the explicit solvent reference to highlight,
while for the implicit solvent structures, we highlight the one with
the lowest RMSD to the explicit solvent reference. It is clear that
for each metastable state, the structures are comparable between systems
with explicit and implicit solvent models. Comparing structures from
the folded and misfolded states (Fig.~\ref{fig:chiStructures}a and
b) with the GBSA\textendash OBC model and with ISSNet models, the
local displacements across the overlayed structures from the latter
are less apparent than from the former. This phenomenon corresponds
to the fact that these states are correctly stabilized by the ISSNet
models (cf.~Fig.~\ref{fig:Results-chi}b and c).

\begin{table}
\caption{\label{tab:Results-chi-metrics}KL divergence, JS divergence and MSE
of free energy for comparing the thermodynamics of the implicit solvent,
vacuum and the explicit solvent systems for chignolin. The metrics
were evaluated based on discrete conformational distributions on the
TIC 1\textendash TIC 2 plane as estimated from the simulation trajectories.
In the case of explicit solvent data set, MSM-reweighting is performed.
Bold font designates the lowest divergence/error values, which correspond
to the implicit solvent model with ISSNet plus charge-only (q-) or
type-and-charge (qt-) embeddings (cf.~Figure~\ref{fig:Results-chi}b
and c).}

\begin{ruledtabular}
\begin{tabular}{lccc}
System & $D_{\textrm{KL}}$ & \multirow{1}{*}{$D_{\textrm{JS}}$} & $\mathrm{MSE}$\footnote{Unit:~(kcal/mol)$^{2}$.}\tabularnewline
\colrule Explicit solvent & (0.) & (0.) & (0.)\tabularnewline
 &  &  & \tabularnewline
t\textendash ISSNet\footnote{The former and latter values on these lines denote the metric values
for corresponding implicit solvent systems with two and three interaction
blocks in the SchNet architecture (see~Section~\ref{subsec:ISSNet-architecture}),
respectively.} & 2.671/0.494 & 0.724/0.117 & 4.438/1.017\tabularnewline
\textbf{q\textendash ISSNet}$^{\textrm{a}}$ & 0.221/0.366 & 0.053/0.086 & \textbf{0.432}/0.526\tabularnewline
\textbf{qt\textendash ISSNet}$^{\textrm{a}}$ & \textbf{0.069}/0.321 & \textbf{0.016}/0.076 & 0.468/0.541\tabularnewline
 &  &  & \tabularnewline
GBSA\textendash OBC & 1.720 & 0.404 & 0.892\tabularnewline
 &  &  & \tabularnewline
Vacuum & 2.647 & 0.726 & 1.815\tabularnewline
\end{tabular}
\end{ruledtabular}

\end{table}

We quantified the comparisons between the conformational distributions
of the explicit solvent systems and the different implicit solvent
models with the criteria introduced for 2D free energy surfaces (see
Section~C in the SI). Table~\ref{tab:Results-chi-metrics} shows
that implicit solvent simulations with the ISSNet models result in
lower divergences/errors with respect to the reference explicit solvent
model comparing to the one with the GBSA\textendash OBC model, indicating
that ISSNet can better reproduce the thermodynamics of a solvated
chignolin system.

As for the effect of hyperparameter choices, we examined the CV\textendash FM
error and the quantified differences in the free energy surfaces (Table~\ref{tab:Results-chi-metrics}).
The parameters that lead to significant differences are the number
of interaction blocks and the embedding strategies. Although adding
a third interaction block to the models generally results with comparable
or even smaller CV\textendash FM errors (see Table~S3 in the SI),
except for the type-embedding ISSNet, this change does not improve
the accuracy according to the three metrics. (This observation is
contradictory to the claim of Ref.~\onlinecite{Wang_ACSCentSci2019}.)
When other hyperparameters are held constant, using the partial charge
embedding (q-) alone results in the lowest MSE of free energy, but
mixed embedding (qt-) leads to the best results according to the two
divergence criteria.

One of the major discrepancies in the implicit solvation methods in
Fig.~\ref{fig:Results-chi} is the relative population of the metastable
states. Especially in the GBSA\textendash OBC case, the unfolded state
of chignolin is over-stabilized. We hypothesize that this behavior
is mainly caused by an inaccurately predicted melting temperature
$T_{m}$, which is the temperature at which the molecule is found
to be folded or unfolded with equal probability in equilibrium.\citep{Tanford_AdvProteinChem1970,Brandts_JAmChemSoc1964}

\begin{table}
\caption{\label{tab:Results-chi-folding-Tm}Estimated folding $T_{m}$ of chignolin
with different solvent models in MD simulations and experimental reference
value.}

\begin{ruledtabular}
\begin{tabular}{lc}
Solvation model for simulation & $T_{m}$ / K\tabularnewline
\colrule Explicit solvent~\citep{Lindorff_Science2011} & 381(361\textendash 393)\tabularnewline
 & \tabularnewline
q\textendash ISSNet\footnote{Model with 2 interaction blocks.} & $\sim$368/$\sim$370\footnote{The former and latter numbers are estimated by assuming constant enthalpy
and entropy changes or constant heat capacity, respectively. {\small{}See
Section~E of the SI for detail.}}\tabularnewline
qt\textendash ISSNet$^{\textrm{a}}$ & $\sim$355/$\sim$355$^{\textrm{b}}$\tabularnewline
 & \tabularnewline
GBSA\textendash OBC\footnote{Estimated from six replicas at {[}250.0, 274.6, 301.7, 331.4, 364.1,
400.0{]} Kelvin in a PT simulation. Can be compared with results in
Ref.~\onlinecite{Anandakrishnan_BiophysJ2015}.} & $\sim$268/$\sim$269$^{\textrm{b}}$\tabularnewline
 & \tabularnewline
Experimental~\citep{Honda_JACS2008} & $\sim$343\tabularnewline
\end{tabular}
\end{ruledtabular}

\end{table}

\begin{figure}
\centering{}\includegraphics[width=1\linewidth]{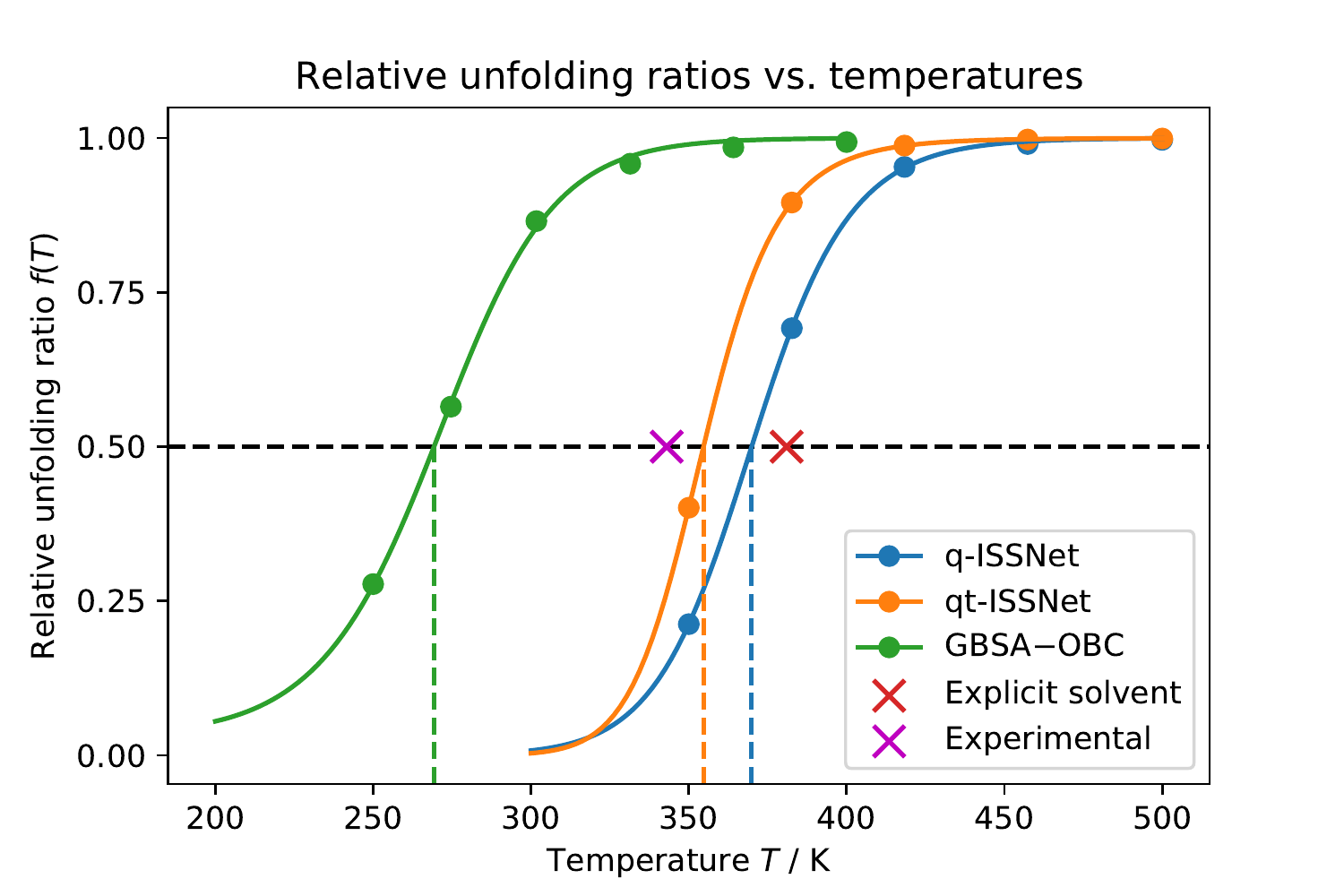}\caption{\label{fig:f(T)}Relative unfolding ratio $f(T)$ for different solvent
models. Here we use the constant-heat-capacity model for curve fitting.
Dashed lines imply the estimated melting temperatures for each cases.
Crosses visualize $T_{m}$s from explicit solvent simulation~\citep{Lindorff_Science2011}
and experiments~\citep{Honda_JACS2008}, which serve as references.}
\end{figure}

The melting temperature is defined as the temperature at which the
molecule is found with equal probability in either the folded or the
unfolded states. This temperature is connected to the zero-crossing
of the unfolding free energy change $\Delta G(T)$, since
\begin{equation}
\Delta G(T)=-\beta^{-1}\log\frac{p_{\textrm{unfolded}}(T)}{p_{\textrm{folded}}(T)}.
\end{equation}
Therefore, we can model the temperature dependency of $\Delta G$
from the sample distributions for the replicas at different temperatures
in the PT simulations, and then solve for $T_{m}$. We utilize the
two models from Ref.~\onlinecite{Schellman_AnnRevBiophysBiophysChem1987}
for $\Delta G-T$ relationship and try to determine the parameters
by curve fitting. It is straightforward to directly work on the $\Delta G-T$
plot, but the $\Delta G$ estimation from the simulations has too
large uncertainty when either $p_{\textrm{unfolded}}$ or $p_{\textrm{folded}}$
is too low. Instead, we calculate and plot the relative unfolding
ratio from the raw data (i.e., the solid dots in Figure~\ref{fig:f(T)}):
\begin{equation}
f(T)=\frac{p_{\textrm{unfolded}}}{p_{\textrm{folded}}+p_{\textrm{unfolded}}}=\frac{1}{1+\exp\left[\beta\Delta G(T)\right]},\label{eq:relativeUnfoldingRatio}
\end{equation}
and estimate the model parameters by a least-square curve fitting.
The resulted $\Delta G-T$ models give a the solid curves in Figure~\ref{fig:f(T)}),
which matches the observations from the raw data. Then we calculated
the temperature corresponding to $\Delta G=0$ (i.e., when the curves
cross the $f(T)=0.5$ line in Figure~\ref{fig:f(T)}) as an estimation
of $T_{m}$ (see Section~E of the SI for details). The resulting
$T_{m}$ for implicit solvent simulation with the ISSNet models and
with the GBSA model are listed in Table~\ref{tab:Results-chi-folding-Tm}.
We also include a reference $T_{m}$ for explicit solvent simulation
with the same force field and water model from Ref.~\onlinecite{Lindorff_Science2011}
(calculated with a different approach; details in SI, Section~E).
This analysis shows that the traditional GBSA model dramatically underestimates
the $T_{m}$, while our neural network ISSNet models result in rather
accurate melting temperatures that are bracketed by the explicit solvent
and experimental observations (labeled in Figure~\ref{fig:f(T)}
as crosses). Note that our models were fitted at one single temperature
and can thus not \emph{generally} expected to make quantitative predictions
at other temperatures. However, the good match observed in this case
is a piece of evidence that the ISSNet method can learn the qualitatively
correct physics.

\section{Discussion\label{sec:Discussion}}

Here we provide some physical interpretation for some choices in our
implementation and experiments and discuss remaining challenges that
call for further investigations.

We leverage an enhanced sampling method for the estimate of the free
energy landscape for chignolin simulation with trained ISSNet models.
Although chignolin is usually regarded as a ``fast-folder'',\citep{Davis_JACS2012,Honda_JACS2008}
transitions among the metastable states, e.g., between the folded
and unfolded states, are rather slow comparing to our simulation timescale.
As a reference, the all-atom explicit solvent folding and unfolding
timescales for chignolin in the NVT\textendash ensemble at 343K is
reported to be 0.6 and 2.4~$\mu$s, respectively,\citep{Lindorff_Science2011}
which are several times longer than our simulation time. In fact,
the generation of our explicit solvent reference data set was also
obtained by means of an enhanced sampling method,\citep{Wang_ACSCentSci2019}
and we reweighted the data set according to a MSM analysis in order
to gain the ground truth of the Boltzmann distribution. For assessing
implicit solvent models, we use the PT\textendash MD to enable a rather
accurate equilibrium sampling within short simulation time, as it
speeds up the state transitions without modifying the thermodynamics
at equilibrium.\citep{Trebst_JChemPhys2006,Earl_PhysChemChemPhys2005}

The ISSNet approach employs a (CG)SchNet architecture with slight
modification for expressing the solvation free energy. In both examples
we found that embeddings (q- and qt-) involving partial atomic charge
led to higher accuracy in the recovered thermodynamics than a traditional
embedding (t-) based solely on the identification of the atom type
(see Table~\ref{tab:Results-ala2-metrics} and~\ref{tab:Results-chi-metrics}).
This result underscores the importance of including electrostatic
information in the network for accurate solvent modeling. It is known
that electrostatic interactions are vital for modeling solvent effects
for both explicit and implicit models.\citep{Adcock_ChemRev2006,Onufriev_WaterModels2018,Ren_QRevBiophys2012,Chen_CurrOpinStructBiol2008,Kleinjung_CurrOpinStructBiol2014}
Although partial atomic charges can be learned and predicted by SchNet~\citep{Schutt_JChemPhys2018}
or other networks~\citep{Nebgen_JChemTheoryComput2018,Sifain_JPhysChemLett2018}
from merely the element-type embedding, such predictions tend to require
a deep network with more interaction blocks and a variety of input
molecules. Our results suggest that it is neither accurate nor efficient
for an implicit solvent model to learn the electrostatics from scratch.
We hypothesize that the new atomic embedding strategy may strengthen
the performance and/or reduce the computational cost for some other
SchNet-based molecular machine learning approaches, such as CGSchNet.

Although our ISSNet models appear more accurate than the reference
methods, they are not free of limitations. Regarding the chignolin
results, we observe that the metastable states are not exactly weighted,
and the free energy surface for the misfolded and unfolded metastable
states slightly differ from the reference. In order to tackle these
problems, we experimented with different training setups, such as
training set composition (e.g., distribution of training data on the
space spanned by the first two TICs) and hyperparameters for SchNet
architectures. We observed different simulation outcomes with resulting
models (e.g., Fig.~\ref{fig:Results-chi}~b,~c and Table~\ref{tab:Results-chi-metrics}),
but we do not yet have an ultimate solution to consistently and systematically
improve the accuracy of the free energy landscape.

We note that the CV\textendash FM error is used to assess the models
and to optimize the hyperparameters in both Refs.~\onlinecite{Wang_ACSCentSci2019}
and~\onlinecite{Husic_JChemPhys2020}. In this work, however, we
found that\textemdash at least for the ISSNet models for chignolin\textemdash there
is no strict correspondence between the lowest CV\textendash FM error
and the highest accuracy (e.g., comparing models with different numbers
of interaction blocks and embedding methods for chignolin, see Section
S2 in the SI). We hypothesize that FM error on a limited data set
may fail to assess the global accuracy of free energy surfaces for
complex systems. High-energy regions\textemdash including transition
paths\textemdash constitute only a tiny proportion of the training
and validation data, because their Boltzmann probability is exponentially
lower than those of major energy minima. Therefore, an erroneous prediction
of the mean force in these regions does not strongly affect the overall
FM loss. Nevertheless, it can cause differences in the height of energy
barriers to the metastable states, resulting in an inaccurate relative
free energy difference and thus a wrong weighting of free energy minima.
This hypothesis also has implications on the model training and hyperparameter
optimizations, because both of them rely on only the FM error but
not the energy or distribution weights. In this sense, combining the
variational FM method with alternative CG schemes (e.g., relative
entropy~\citep{Shell_JChemPhys2008,Chaimovich_JChemPhys2011}) may
systematically improve the accuracy of related machine learning methods.

Another aspect to be improved for the ISSNet models is the speed of
simulation (see the SI, Section~F). Because the forces from the neural
network are required for every time step, simulations become computationally
demanding and time-consuming, restricting the application of the current
ISSNet model to longer simulations and larger molecules. In this work
we partially avoided this problem by evaluating the ISSNet forces
in batch, which speeds up the sampling but not single simulations.
While this work presents an important feasibility study, future developments
will involve reducing the frequency of neural network evaluation (e.g.,
by multiple time-step MD simulation), lowering communication overhead
between the MD software and the deep-learning framework as well as
finding computationally cheaper energy neural networks in substitution
for SchNet.\citep{Schutt_JChemPhys2018}

To illustrate the advantage of the ISSNet approach, we compared it
to GBSA\textendash OBC,\citep{Onufriev_Proteins2004} an existing
widely-used implicit solvent model. This choice is due to the availability
in simulation tools such as AMBER~\citealp{Amber20} and OpenMM.\citep{Eastman_PLoSComputBiol2017}
Additionally, a recent study assures the qualitative similarity between
GBSA\textendash OBC and a newer GBNeck2 model~\citealp{Nguyen_JChemTheoryComput2013}
for the implicit solvation of chignolin (CLN025).\citealp{Anandakrishnan_JChemTheoryComput2018}
However, given the wealth of existing implicit solvent methods, we
can not conclude that the ISSNet models trained herein reflect the
state of the art for the accuracy of thermodynamics. Nevertheless,
due to the variational nature of the formulation, given sufficient
training data and a sufficiently competent neural network, our model
shall be able to reproduce the thermodynamics of a given explicit
solvent model with arbitrarily high accuracy.

Despite its success, an ISSNet model is at the moment only parameterized
for a given molecular system at a fixed thermodynamic state. Even
when a model successfully learns the free energy surface specific
to the given system, it is not guaranteed to output sensible solvation
forces for systems at a different temperature/pressure and/or consisting
of other solute molecules. Although we achieved an accurate estimation
of the unfolding temperature $T_{m}$ by the ISSNet models, it may
merely be due to the fact that the simulation temperature for the
data set generation is close to $T_{m}$. In fact, we observed that
the empirical thermodynamic parameters (e.g., the enthalpy and entropy
changes) from curve fitting for chignolin unfolding in implicit solvents
are different from the experimental and explicit solvent results,
thus leading to a significant deviation of the folded population at
other temperatures (see the SI, Section~E). Therefore, a proper modeling
of the temperature/pressure dependence of the free energy surface
is yet to be developed.

Another potential of the future development of the ISSNet method is
to achieve the transferability among a larger variety of solute molecules.
Since the (CG)SchNet architecture allows the same set of parameters
to be shared among models for different systems,\citep{Husic_JChemPhys2020,Schutt_JChemPhys2018}
it is in principle feasible to optimize ISSNet models for a more general
description of the solvent effects. Note that a variety of systems
may also provide information for correctly treating the conformations
that are under-sampled in case of a single training system, thus beneficial
to the accuracy in free-energy modeling at the same time. By training
on extended data sets (e.g., a set of peptides or proteins) and potentially
incorporating more insights from statistical physics, we may train
more \emph{transferable} yet accurate solvation models and widen the
application of the ISSNet approach.

\section{Conclusions}

In this work, we have reformulated the implicit solvation modeling
as a bottom-up coarse graining problem, and shown that an accurate
implicit solvent model can be machine-learned by leveraging the variational
FM approach. Based on the CGnet~\citep{Wang_ACSCentSci2019} and
CGSchNet~\citep{Husic_JChemPhys2020} methods established for machine
learning of CG potentials, we develop ISSNet for learning an implicit
solvent model from explicit solvent simulation data. Our method outperforms
the GBSA\textendash OBC model~\citep{Onufriev_Proteins2004}\textemdash an
widely used implicit solvent method\textemdash on two biomolecular
benchmark systems (capped alanine and chignolin) in terms of accuracy.
Our novel method sets up a stage for utilizing the power of machine
learning to the implicit solvent problem, and we expect further development
on the transferability among thermodynamic states and chemical space
to widen its application.

\section*{Supporting Information}

Detailed setups for model training and simulation, as well as procedures
for various analyses that are referred to in the main text can be
found in the online supplementary material.
\begin{acknowledgments}
The authors would like to thank Adrià Pérez and Gianni de Fabritiis
for providing the chignolin data set and details about their setup,
Simon Olsson, Tim Hempel, Moritz Hoffmann, Dr.~Jan Hermann, Zeno
Schätzle and Jonas Köhler for insightful discussions on molecular
dynamics and/or machine learning. Y.C., A.K., B.E.H.~and F.N.~gratefully
acknowledge funding from European Commission (Grant No.~ERC CoG 772230
\textquotedblleft ScaleCell\textquotedblright ), the International
Max Planck Research School for Biology and Computation (IMPRS\textendash BAC),
the BMBF (Berlin Institute for Learning and Data, BIFOLD), the Berlin
Mathematics center MATH+~(AA1-6, EF1-2) and the Deutsche Forschungsgemeinschaft
DFG (SFB1114/A04). N.E.C.~and C.C.~acknowledge National Science
Foundation (CHE-1738990, CHE-1900374, and PHY-2019745), the Welch
Foundation (C-1570), the Deutsche Forschungsgemeinschaft (SFB/TRR
186/A12, and SFB 1078/C7), and the Einstein Foundation Berlin. The
3D molecular structures are visualized with PyMOL~\citep{PyMOL}.
\end{acknowledgments}

\section*{Data Availability}

The data that support the findings of this study are available from
the corresponding author upon reasonable request.


%\bibliographystyle{aipnum4-1}
%\bibliography{bibtex/ala2,bibtex/chi,bibtex/intro,bibtex/methods,bibtex/theory}

\end{document}